\documentclass[slac_one]{revtex4}
\usepackage{graphicx,fancyhdr,amsmath,axodraw}
\allowdisplaybreaks

\newcommand{\eg}{e.g.\ }
\newcommand{\ie}{i.e.\ }
\newcommand{\M}{\mathcal{M}}
\newcommand{\ri}{\mathrm{i}}
\newcommand{\ket}[1]{\left| #1\right\rangle}
\newcommand{\bra}[1]{\left\langle #1\right|}
\newcommand{\Tr}{\mathop{\mathrm{Tr}}}
\newcommand{\uscore}{\symbol{95}}
\newcommand{\lbrac}{\symbol{123}}
\newcommand{\rbrac}{\symbol{125}}

\begin{document}

\title{{\small{2005 International Linear Collider Workshop - Stanford,
U.S.A.}}\\ 
\vspace{12pt}
New Developments in FormCalc 4.1} 

\author{Thomas Hahn}
\affiliation{Max-Planck-Institut f\"ur Physik,
F\"ohringer Ring 6,
D--80805 Munich, Germany}

\begin{abstract}
FormCalc is a matrix-element generator that turns FeynArts amplitudes up 
to one loop into a Fortran code for computing the squared matrix element.
The generated code can be run with FormCalc's own driver programs or 
used  with other `frontends', \eg Monte Carlos.  Major new or enhanced 
features in Version 4.1 are: treatment of external fermions, phase-space 
integration, code-generation functions, extensions for the MSSM, the 
HadCalc frontend.
\end{abstract}

\maketitle

\thispagestyle{fancy}


\section{Introduction}

FormCalc \cite{FormCalc} is a package for the calculation of Feynman
amplitudes based on Mathematica and FORM \cite{FORM}.  Amplitudes
generated by FeynArts \cite{FeynArts} are first simplified analytically
and then converted to a self-contained Fortran code for the computation
of the squared matrix element.  Currently, diagrams up to one loop can
be simplified, and kinematics are supplied for $1\to 2$, $2\to 2$, and
$2\to 3$ processes.

The present article describes the features added or enhanced in 
version 4.1.  They fall into several categories:
\begin{itemize}
\itemsep=0pt
\item Weyl--van der Waerden (WvdW) spinor formalism for external
      fermion lines,
\item Multidimensional phase-space integration with Cuba 1.2,
\item Public functions for writing out Mathematica expressions as
      optimized Fortran code,
\item FeynHiggs interface,
\item Non-minimal flavour violation (NMFV) for the MSSM,
\item Various useful scripts,
\item The HadCalc frontend, by M.~Rauch, for hadronic calculations.
\end{itemize}


\section{Weyl--van der Waerden spinor formalism}

Amplitudes involving external fermions have the form $\M = \sum_{i =
1}^n c_i\, F_i$, where the $F_i$ are (products of) fermion chains.  The
textbook recipe is to compute probabilities, such as $|\M|^2 = \sum_{i,
j = 1}^n c_i^*\, c_j\, F_i^* F_j$, and evaluate the $F_i^* F_j$ by
standard trace techniques: $\left|\bra{u}\Gamma\ket{v}\right|^2
= \bra{u}\Gamma\ket{v}\bra{v}\bar\Gamma\ket{u} 
= \Tr\bigl(\Gamma\ket{v}\bra{v}\bar\Gamma\ket{u}\bra{u}\bigr)$.

The problem with this approach is that instead of $n$ of the $F_i$ one
needs to compute $n^2$ of the $F_i^* F_j$.  Since essentially $n\sim
(\text{number of vectors})!$, this quickly becomes a limiting factor in
problems involving many vectors, \eg in multi-particle final states or
polarization effects.

The solution is of course to compute the amplitude $\M$ directly and
this is done most conveniently in the WvdW formalism \cite{WvdW}.  The
implementation of this technique in an automated program has been
outlined in \cite{Optimizations}.

The \texttt{FermionChains} option of \texttt{CalcFeynAmp} determines how
fermion chains are returned: \texttt{Weyl}, the default, selects Weyl
chains.  \texttt{Chiral} and \texttt{VA} select Dirac chains in the
chiral ($\omega_+/\omega_-$) and vector/axial-vector ($1/\gamma_5$)
decomposition, respectively.  The Weyl chains do not need to be further
evaluated with \texttt{HelicityME}, which applies the trace technique.

The WvdW method has other advantages, too: Polarization does not `cost'
extra in terms of CPU time, that is, one gets the spin physics for free. 
Whereas with the trace technique the formulas become significantly more
bloated when polarization is taken into account, in the WvdW formalism
one actually needs to sum up the polarized amplitudes to get the
unpolarized result.  There is also better numerical stability because 
components of $k^\mu$ are arranged as `large' and `small' matrix 
entries, viz.
\begin{equation}
\sigma_\mu k^\mu = \begin{pmatrix}
k_0 + k_3 & k_1 - \ri k_2 \\
k_1 + \ri k_2 & k_0 - k_3
\end{pmatrix}.
\end{equation}
Cancellations of the form $k_0 - k = \sqrt{k^2 + m^2} - \sqrt{k^2}$ for
$m\ll k$ can be avoided (\eg by observing that $k_0 - k = m^2/(k_0 + k)$ 
for on-shell $k$) and hence mass effects are treated more accurately.


\section{Multidimensional integration with Cuba}

FormCalc uses the Cuba library \cite{Cuba} for numerical integration of
multidimensional phase-spaces.  Cuba provides four integration routines:
Vegas, Suave, Divonne, and Cuhre.  All four have a very similar
invocation and can thus be interchanged easily, \eg for comparison.  The
flexibility of a general-purpose method is particularly useful in the
setting of automatically generated code.  The following table gives an
overview of the features of the Cuba routines; specific details on the
implementation and the actual usage of the Cuba routines are provided in
\cite{Cuba} and shall not be repeated here.
\begin{center}
\begin{tabular}{llllllll}
Routine  &\hspace*{1em}&
	Basic method &\hspace*{1em}&
	Type &\hspace*{1em}&
	Variance reduction \\
\hline
Vegas &&
	Sobol sample &&
	quasi-Monte Carlo &&
	importance sampling \\
&&
	\textit{or} Mersenne Twister sample &&
	pseudo-Monte Carlo && \\
\hline
Suave &&
	Sobol sample &&
	quasi-Monte Carlo &&
	globally adaptive subdivision \\
&&
	\textit{or} Mersenne Twister sample &&
	pseudo-Monte Carlo &&
	\quad + importance sampling \\
\hline
Divonne &&
	Korobov sample &&
	lattice method &&
	stratified sampling, \\
&&
	\textit{or} Sobol sample &&
	quasi-Monte Carlo &&
	\quad aided by methods from \\
&&
	\textit{or} Mersenne Twister sample &&
	pseudo-Monte Carlo &&
	\quad numerical optimization \\
&&
	\textit{or} cubature rules &&
	deterministic \\
\hline
Cuhre &&
	cubature rules &&
	deterministic &&
	globally adaptive subdivision \\
\hline
\end{tabular}
\end{center}

\begin{figure}
\centerline{\includegraphics[width=.5\hsize]{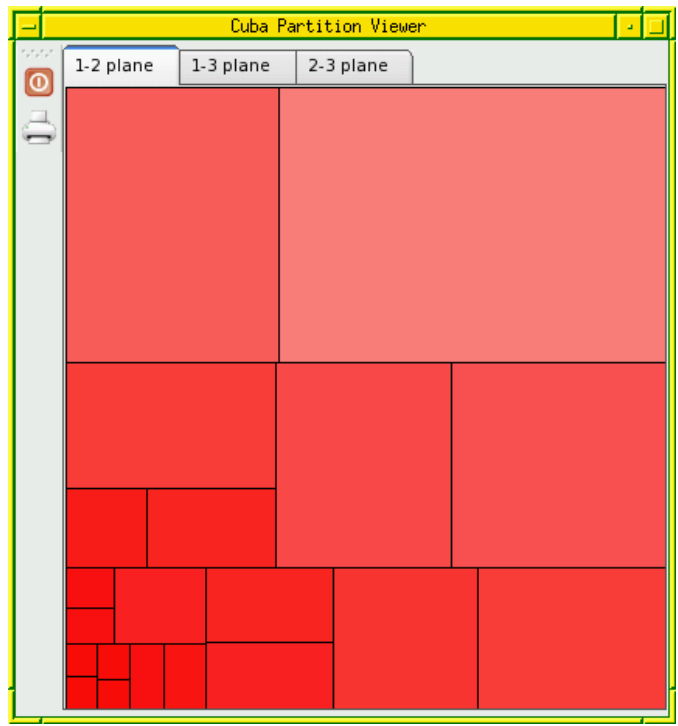}}
\caption{\label{fig:partview}A screenshot of the Cuba Partition Viewer.}
\end{figure}

Apart from some important bug-fixes, Cuba Version 1.2 includes the 
following new features:
\begin{itemize}
\item
Pseudo-random sampling using the Mersenne Twister generator has been 
added to all Monte Carlo algorithms.  This is useful not only for 
comparison, but also because in high dimensions the numerator in the 
convergence rate for quasi-random samples, $\mathcal{O}(\log^{d - 1} 
n/n)$, becomes noticeable.

\item
Vegas can memorize its internal grid for subsequent invocations, to 
speed up integration on similar integrands.

\item
Vegas can save its internal state in a file such that the calculation 
can be resumed \eg after a crash.

\item
There exists a one-stop invocation for the Cuba routines, \\
\hspace*{2em}\texttt{subroutine Cuba(ndim, integrand, result, error)} \\
with which the number of parameters that need to be passed is reduced to
a minimum.  (Note that this particularly simple form is special to 
FormCalc, and that Cuba's original definition is slightly more 
involved.)

\item
A partition viewer has been added.  It allows to view the partitioning
of the integration region performed by Suave, Divonne, and Cuhre (Vegas 
does not tessellate the integration region).  To use the partition 
viewer, set the verbosity level to 3 and pipe the output of your program 
through the \texttt{partview} program, as in \\
\hspace*{2em}\texttt{myprogram | partview 1 2 1 3} \\
Each pair of numbers on the command-line refers to a hyperplane to 
display, \ie in the above example the 1--2 and 1--3 hyperplanes are 
shown.  See Fig.\ \ref{fig:partview} for a screenshot.
\end{itemize}

Cuba possesses also a Mathematica interface.  The MathLink executables
are loaded \eg with \texttt{Install["Vegas"]} and make the Cuba routines
available in a way almost like
\texttt{NIntegrate}.
\begin{center}
\begin{picture}(300,90)
\Text(50,85)[b]{Mathematica}
\CBox(0,0)(100,80){Red}{PastelRed}
\Text(50,65)[]{\tt Vegas[$f$, $\dots$]}
\Text(50,20)[]{integrand $f$}

\SetOffset(200,0)
\Text(50,85)[b]{C}
\CBox(0,0)(100,80){Blue}{PastelBlue}
\Text(50,65)[]{{\tt void Vegas($\dots$)}}
\Text(50,20)[]{request samples}

\SetWidth{.5}
\SetOffset(0,0)
\LongArrow(50,55)(50,31)
\CBox(33.5,39.5)(66.5,49){Black}{PastelRed}
\Text(50,44)[]{{\scriptsize\tt Compile}}

\SetWidth{2}
\SetOffset(150,0)
\LongArrow(-40,65)(40,65)
\Text(0,70)[b]{MathLink}

\Text(0,28)[b]{$\{\vec x_1, \vec x_2, \dots\}$}
\LongArrow(40,25)(-40,25)
\LongArrow(-40,15)(40,15)
\Text(0,12)[t]{$\{f_1, f_2, \dots\}$}
\end{picture}
\end{center}
The integrand stays in Mathematica at all times (see figure above), 
which means that one can integrate functions which are not simple to 
implement in C or Fortran, \eg
\texttt{Cuhre[Zeta[x y], \lbrac x,.2,.3\rbrac, \lbrac y,.4,.5\rbrac]}.


\section{Public code-generation functions}

FormCalc's code-generation functions, hitherto used only internally,
have been reorganized for greater user-friendliness and are now publicly
available.  They can be used to write out an arbitrary Mathematica
expression as optimized Fortran code.  The basic procedure is very
simple:
\begin{enumerate}
\itemsep=0pt
\item
{\tt {\it handle} = OpenFortran["{\it file.F}"]} \\
opens {\it file.F} as a Fortran file for writing,   

\item
{\tt WriteExpr[{\it handle}, \lbrac {\it var} ->
  {\it expr}, \dots \rbrac]} \\
writes out Fortran code to calculate {\it expr} and store the
result in {\it var},

\item
{\tt Close[{\it handle}]} \\
closes the file again.
\end{enumerate}
The code generation is fairly sophisticated and goes well beyond
merely applying Mathematica's \texttt{FortranForm}.  The generated code 
is optimized, \eg common subexpressions are pulled out and computed in 
temporary variables.  Expressions too large for Fortran are split into 
parts, as in
\begin{verbatim}
     var = part1
     var = var + part2
     ...
\end{verbatim}
If the expression is too large even to be sensibly evaluated in one 
file, the \texttt{FileSplit} function can distribute it on several 
files and optionally write out a master subroutine which calls the 
individual parts.

To further automate the code generation, such that the resulting code 
needs few or no changes by hand, many ancillary functions are available, 
\eg \texttt{CommonDecl} writes out common-block declarations for a given 
list of variables.


\section{FeynHiggs interface and NMFV}

FormCalc's initialization code for the MSSM, \texttt{mssm\uscore ini.F},
now contains an interface to FeynHiggs \cite{FeynHiggs,FeynHiggs2.2} for
the computation of the Higgs masses.  Since in particular the light
Higgs-boson mass receives significant radiative corrections, its precise
determination is phenomenologically important.  From the user
perspective, a preprocessor flag governs the choice of Higgs masses:
\begin{itemize}
\itemsep=0pt
\item
\texttt{\#define HIGGS\uscore MASSES TREE} \\
use the tree-level Higgs masses,

\item
\texttt{\#define HIGGS\uscore MASSES SIMPLE} \\
use a simple one-loop formula,

\item
\texttt{\#define HIGGS\uscore MASSES FEYNHIGGS} \\
invoke FeynHiggs 2.2 \cite{FeynHiggs2.2}
-- this is the most precise determination,

\item
\texttt{HIGGS\uscore MASSES} undefined \\
use the FeynHiggsFast approximation \cite{FeynHiggsFast}
-- quite precise, but valid only for real SUSY parameters.
\end{itemize}

Also for the MSSM, non-minimal flavour violation (NMFV) can be enabled. 
In the usual, minimal, setup, there is no flavour-violation beyond the
Standard Model's CKM effects, \ie left--right mixing is independent for
each squark flavour.  In contrast, NMFV means full $6\times 6$ mixing
among both up-type squarks $(\tilde u_1, \tilde u_2, \tilde c_1, \tilde
c_2, \tilde t_1, \tilde t_2)$ and down-type squarks $(\tilde d_1, \tilde
d_2, \tilde s_1, \tilde s_2, \tilde b_1, \tilde b_2)$, see \eg 
\cite{NMFV}.

On the technical side, the diagrams have to be generated using the
\texttt{FVMSSM.mod} model file in FeynArts, and in the numerical
evaluation the preprocessor flag \texttt{FLAVOUR\uscore VIOLATION} has
to be defined.  Once this flag is enabled, the $6\times 6$ mass matrix
is accessible as \texttt{LambdaSf} and in addition to the usual squark
masses and mixing matrices, \texttt{MSf} and \texttt{USf}, there exist
\texttt{MASf} and \texttt{UASf} containing the corresponding NMFV
quantities.


\section{Shell scripts}

FormCalc 4.1 includes a few useful shell scripts:
\begin{itemize}
\item
\texttt{sfx} packs all source files in the directory it is invoked in 
into a mail-safe self-extracting archive.  The archived code in 
particular does not need FormCalc to compile or run.

\item
\texttt{turnoff} switches off (and on) the evaluation of certain parts
of the amplitude, which is a handy thing for testing.  For example,
``\texttt{turnoff box}'' turns off the boxes (actually, all parts of the
amplitude with `box' in their name).  \texttt{turnoff} without arguments
then restores all modules.

\item
\texttt{pnuglot} produces a high-quality plot in Encapsulated PostScript
format from a data file in just one line.

\item
\texttt{submit} automatically distributes a parameter scan on a cluster.  
The available machines first have to be declared in a file 
\texttt{.submitrc}, \eg
\begin{verbatim}
    # Optional: `nice' to start jobs with
    nice 10
    # Pentium 4
    pcl301
    pcl305
    # Dual Xeon
    pcl247b   2
    pcl319a   2
    ...
\end{verbatim}
After that the command line, typically something like
``\texttt{run uuuu 500,1000}'', simply has to be prefixed by 
\texttt{submit}, \ie ``\texttt{submit run uuuu 500,1000}''.
\end{itemize}


\section{The HadCalc frontend}

HadCalc is a new frontend for FormCalc, \ie it uses the generated
Fortran code with a custom set of driver programs.  It was written by
Michael Rauch who also maintains it independently from FormCalc.

HadCalc automates the calculation of hadronic cross-sections.  It
automatically performs the convolution with PDFs and allows various cuts
to be applied.  Cross-sections can be computed either fully integrated
or differential in invariant mass, rapidity, or transverse momentum. 
HadCalc operates either in batch mode, like FormCalc, or interactively,
which allows the user \eg to play with parameters.

The program is not (yet) public.  It can currently be obtained on 
request from $\langle$mrauch@mppmu.mpg.de$\rangle$.


\section{Summary}

FormCalc 4.1 is the current release of the FormCalc package.  It has
many new features enhancing performance and user-friendliness.  The new
public Fortran code-generation functions should be interesting also to
people who do not use FormCalc to compute Feynman amplitudes.  The
package is available as open source and stands under the GNU Lesser
General Public License (LGPL).  It can be obtained from the Web site
\texttt{http://www.feynarts.de/formcalc}.


\newcommand{\volyearpage}[3]{\textbf{#1} (#2) #3}
\newcommand{\cpc}{\textsl{Comp.\ Phys.\ Commun.} \volyearpage}
\newcommand{\pr}{\textsl{Phys.\ Rev.} \volyearpage}
\newcommand{\npps}{\textsl{Nucl.\ Phys.\ Proc.\ Suppl.} \volyearpage}

\begin{flushleft}

\end{flushleft}

\end{document}